\documentclass[aps, pra, twocolumn, amsmath, amssymb, superscriptaddress, nofootinbib]{revtex4-1}

\makeatletter
\def\p@subsection{}
\def\p@subsubsection{}
\makeatother

\usepackage[utf8]{inputenc}
\usepackage{graphicx}
\usepackage{units}
\usepackage{color}
\usepackage[pdftex, colorlinks=true, linkcolor=myblue, citecolor=myblue,
urlcolor=myblue]{hyperref}
\usepackage{times}
\usepackage{comment}
\usepackage{graphicx}
\usepackage{amsmath, calc}
\usepackage{mathrsfs}
\usepackage{amsfonts}
\usepackage{amssymb}
\usepackage{bm}
\usepackage{bbm}
\usepackage{color}
\usepackage{array}
\usepackage{units}
\usepackage{tabstackengine}
\stackMath

\definecolor{myblue}{rgb}{0,0,1}
\definecolor{myred}{rgb}{1,0,0}

\DeclareMathOperator{\Tr}{Tr}

\DeclareMathOperator\arctanh{arctanh}
\DeclareMathOperator\arccot{arccot}
\DeclareMathOperator\arccoth{arccoth}

\begin{document}


\title{Hyperbolic enhancement of a quantum battery}


\author{C.  A. Downing} 
\email{c.a.downing@exeter.ac.uk}
\affiliation{Department of Physics and Astronomy, University of Exeter, Exeter EX4 4QL, United Kingdom}

\author{M. S. Ukhtary}
\affiliation{Department of Physics and Astronomy, University of Exeter, Exeter EX4 4QL, United Kingdom}
\affiliation{Research Center for Quantum Physics, National Research and Innovation Agency (BRIN), South Tangerang 15314, Indonesia}


\date{\today}


\begin{abstract}
A quantum system which can store energy, and from which one can extract useful work, is known as a quantum battery. Such a device raises interesting issues surrounding how quantum physics can provide certain advantages in the charging, energy storage or discharging of the quantum battery as compared to their classical equivalents. However, the pernicious effect of dissipation degrades the performance of any realistic battery. Here we show how one can circumvent this problem of energy loss by proposing a quantum battery model which benefits from quantum squeezing. Namely, charging the battery quadratically with a short temporal pulse induces a hyperbolic enhancement in the stored energy, such that the dissipation present becomes essentially negligible in comparison. Furthermore, we show that when the driving is strong enough the useful work which can be extracted from the quantum battery, that is the ergotropy, is exactly equal to the stored energy. These impressive properties imply a highly efficient quantum energetic device with abundant amounts of ergotropy. Our theoretical results suggest a possible route to realizing high-performance quantum batteries, which could be realized in a variety of platforms exploiting quantum continuous variables.
\end{abstract}


\maketitle



\noindent \textbf{Introduction}\\
The emerging field of quantum energy science seeks to discover unprecedented improvements in an array of promising energy technologies, from quantum solar power to nuclear fusion~\cite{Metzler2023}. Amongst these innovations, the necessity to improve the world's existing energy storage technology ensures that quantum battery research is of utmost importance. Interestingly, as well as quantum physics and engineering, quantum battery research encompasses both quantum thermodynamics and quantum information science, such that this subfield poses appealing fundamental and applied challenges~\cite{Campaioli2023}.

Early and inventive experiments probing the transfer and storage of energy in the quantum world have already revealed certain quantum advantages. During the charging process, superextensive charging rates have been reported in an organic microcavity~\cite{Quach2022}, quantum correlations-boosted charging has been demonstrated in a spin system~\cite{Joshi2022}, and the manipulation of bright and dark states has led to stable charging in a superconducting quantum battery~\cite{Hu2022}. Meanwhile, careful measurements of the energy transfer between light and qubits has benchmarked the current capabilities for controlling quantum energy and information flow~\cite{Stevens2022, Maillette2022}.

\begin{figure}[tb]
 \includegraphics[width=\linewidth]{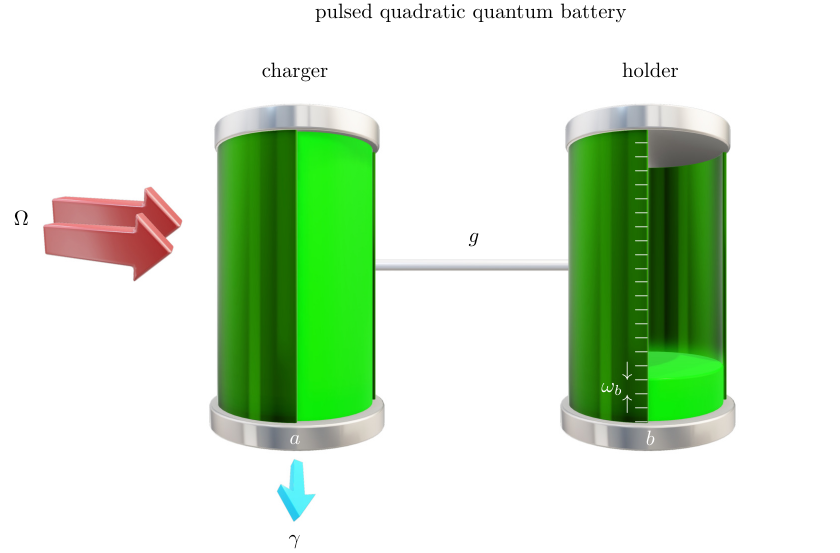}
 \caption{ \textbf{The pulsed quadratic quantum battery.} A representation of the considered bipartite quantum battery model of Eq.~\eqref{eq:Haxcsdfdsfdsfsdfvcxvmy}, associated with the two bosonic operators $a$ and $b$ [cf. Eq.~\eqref{eq:dgvgbbbbb} and Eq.~\eqref{eq:sfdvvsssc}]. The battery charger is coupled (grey bar) at the rate $g$ to the battery holder [cf. Eq.~\eqref{eq:vdsfsdcsss}], which has the level spacing $\omega_b$ (white lines). The charger is driven by a quadratic pulse (red arrows) of dimensionless strength $\Omega$ [cf. Eq.~\eqref{eq:sfdcvfdddn}], and suffers from dissipation (cyan arrow) at the decay rate $\gamma$ [cf. Eq.~\eqref{eq:dfdfdf}]. In this sketch, there is an example stored energy $E = 3 \omega_b$ in the battery holder [cf. Eq.~\eqref{eq:fgdgfgd}].}
 \label{figure0}
\end{figure}

However, some problems which harm the expected utility of quantum batteries remain. Dissipation is intrinsic to open quantum systems, so how to effectively combat energy loss from a quantum battery? Thermodynamically, there is a maximum amount of work which can be reasonably extracted from a quantum storage device (the so-called ergotropy)~\cite{Allahverdyan2004, Alicki2013}, so how to optimize this measure instead of the overall stored energy? A stream of theoretical proposals for model quantum batteries have sought to address these important issues, as well as related questions concerning the speed-up of the charging and discharging processes~\cite{Andolina2018, Friis2018, Farina2019, Santos2020, NewCrescente2020, Santos2021, Shaghaghi2002, Gyhm2022,Konar2022, Mazzoncini2023, Centrone2021, Catalano2023, Song2024, Ukhtary2024}. Here we suggest a rather different approach based upon quantum squeezing. We consider a bipartite quantum battery model, where the battery charger is coupled to the battery holder [cf. the sketch of Fig.~\ref{figure0}]. Importantly, we propose that the charger is driven by a quadratic~\cite{Savona2017, Rota2019, Garbe2020,Campinas2023} temporal pulse, which induces squeezing into the system. Consequentially, there is a hyperbolic enhancement in the energy stored in the battery holder, such that the inevitable energy loss in the system is made effectively negligible. Previously, continuous (rather than pulsed) two-photon drives have been studied theoretically within the context of quantum batteries in Ref.~\cite{Sassetti2020, Downing2023, Fan2023}.

The total Hamiltonian operator $\hat{H}$ of the composite quantum battery system, as sketched in Fig.~\ref{figure0}, can be decomposed into four parts,
\begin{equation}
\label{eq:Haxcsdfdsfdsfsdfvcxvmy}
 \hat{H} =  \hat{H}_a +  \hat{H}_b +  \hat{H}_{a-b} +  \hat{H}_{d},
\end{equation}
which account for the battery charger and holder energies, the charger-holding coupling, and the driving of the battery charger as follows (we take $\hbar = 1$ here and throughout)
\begin{align}
 \hat{H}_a &= \omega_b \, a^\dagger a, \label{eq:dgvgbbbbb} \\
  \hat{H}_b &= \omega_b \, b^\dagger b, \label{eq:sfdvvsssc} \\
 \hat{H}_{a-b} &= g  \left( a^\dagger b + b^\dagger a \right), \label{eq:vdsfsdcsss} \\
   \hat{H}_d &= \tfrac{\Omega}{2} \, \delta(t) \left( a^\dagger a^\dagger + a a \right). \label{eq:sfdcvfdddn}
\end{align}
The bosonic creation and annihilation operators $b^\dagger$ and $b$ raise and lower respectively the number of excitations in the battery holder, and they are subject to the commutation rule $[ b, b^\dagger ] = 1$, while $a^\dagger$ and $a$ behave similarly for the battery charger. We consider the energy level spacing $\omega_b$ for the charger and the holder to be the same [cf. Eq.~\eqref{eq:dgvgbbbbb} and Eq.~\eqref{eq:sfdvvsssc}], and there is a coupling rate $g > 0$ between them which mediates the energy transfer [cf. Eq.~\eqref{eq:vdsfsdcsss}]. In turn, the battery charger is driven quadratically with a short temporal pulse of dimensionless strength $\Omega$, and $\delta(x)$ is the delta function of Dirac [cf. Eq.~\eqref{eq:sfdcvfdddn}]. Experimentally, such kinds of parametric (or two-photon) drives have been recently realized using a variety of superconducting circuits~\cite{Leghtas2015, Wang2019, Chien2020, Gaikwad2023}, while quantum harmonic oscillators are known to nicely model realistic quantum batteries~\cite{Qu2023}.

The Hamiltonian dynamics of Eq.~\eqref{eq:Haxcsdfdsfdsfsdfvcxvmy} can be upgraded with the aid of a Gorini–Kossakowski–Sudarshan–Lindblad quantum master equation in order to account for the inevitable energy loss from the battery charger into the external environment. The dissipationless battery holder is assumed to be isolated from the environment, since it is supposed to be able to store energy on an exceedingly large timescale. This approximation is indeed close to the current experimental reality~\cite{Quach2022, Joshi2022, Hu2022, Stevens2022, Maillette2022}. The density matrix $\rho$ of the overall quantum battery is then governed by the master equation~\cite{Breuer2002}
\begin{equation}
\label{eq:dfdfdf}
\partial_t \rho = \mathrm{i} [ \rho,  \hat{H}] + \frac{\gamma}{2} \left( 2 a \rho a^\dagger - a^\dagger a \rho - \rho a^\dagger a \right),
\end{equation}
where the Hamiltonian operator $\hat{H}$ is defined by Eq.~\eqref{eq:Haxcsdfdsfdsfsdfvcxvmy}, and where $\gamma \ge 0$ is the decay rate of the battery charger. This completes the mathematical description of the model, as represented pictorially in Fig.~\ref{figure0}. The average value of the battery holder population $\langle b^\dagger b \rangle$, which can be accessed from Eq.~\eqref{eq:dfdfdf} after employing the trace property $\Tr ( \mathcal{O} \rho ) = \langle  \mathcal{O}  \rangle$ with the density matrix $\rho$ (where $\mathcal{O}$ is some operator), then gives rise to the energy $E$ stored in the quantum battery 
\begin{equation}
\label{eq:fgdgfgd}
 E = \omega_b \langle b^\dagger b \rangle. 
\end{equation}
Determining this important energetic measure, which takes on the example value of $E = 3\omega_b$ in the cartoon battery imagined in Fig.~\ref{figure0}, for the introduced driven-dissipative model of Eq.~\eqref{eq:dfdfdf} is a key aim of this theoretical study, alongside finding the optimal time needed to achieve the energetic maximum. 

After the battery charger is disconnected from the battery holder, the dissipationless holder is then able to store the energy $E$ indefinitely. However, thermodynamically not all of this stockpiled energy is useful for doing work. The useful energy stored in the quantum battery, or more formally the so-called ergotropy $\mathcal{E}$, is given by the celebrated formula~\cite{Allahverdyan2004, Alicki2013}
\begin{equation}
\label{eq:fgddsfdfsfgfgd}
 \mathcal{E} =  E -  E_\beta, 
\end{equation}
which subtracts off from the stored energy $E$ of Eq.~\eqref{eq:fgdgfgd} the unuseful energy $E_\beta$ of the passive state $\rho_\beta$. Within quantum continuous variable theory~\cite{Olivares2012, Serafini2017}, the passive state energy $E_\beta$ can be written down exactly in terms of the second moments $\langle b^\dagger b \rangle$ and $\langle b b \rangle$ of the battery holder, as follows~\cite{Farina2019, Downing2023}
\begin{align}
\label{eq:fgdgljklfgd}
  E_\beta &= \omega_b \left( \tfrac{ \sqrt{\mathcal{D}}-1}{2} \right), \\
    \mathcal{D} &= \left( 1 + 2 \langle b^\dagger b \rangle  \right)^2 - 4 \big| \langle b b \rangle  \big|^2, \label{eq:svgverfeee}
\end{align}
since here all first moments are zero (that is $\langle b \rangle = \langle b^\dagger \rangle = 0$, as is shown in the Supplemental Material~\cite{SupplementalMaterial}). In what follows, we describe analytically the energetic $E$ and ergtropic $\mathcal{E}$ measures of the presented model, revealing the high performing nature of the proposed quadratic quantum battery. We also note that whilst we consider a delta pulse in Eq.~\eqref{eq:sfdcvfdddn}, our forthcoming analytic results are also applicable to driving pulses with some finite temporal width $\tau$, as long as the timescale $\omega_b \tau \ll 1$ is observed.
\\

\begin{figure*}[tb]
 \includegraphics[width=\linewidth]{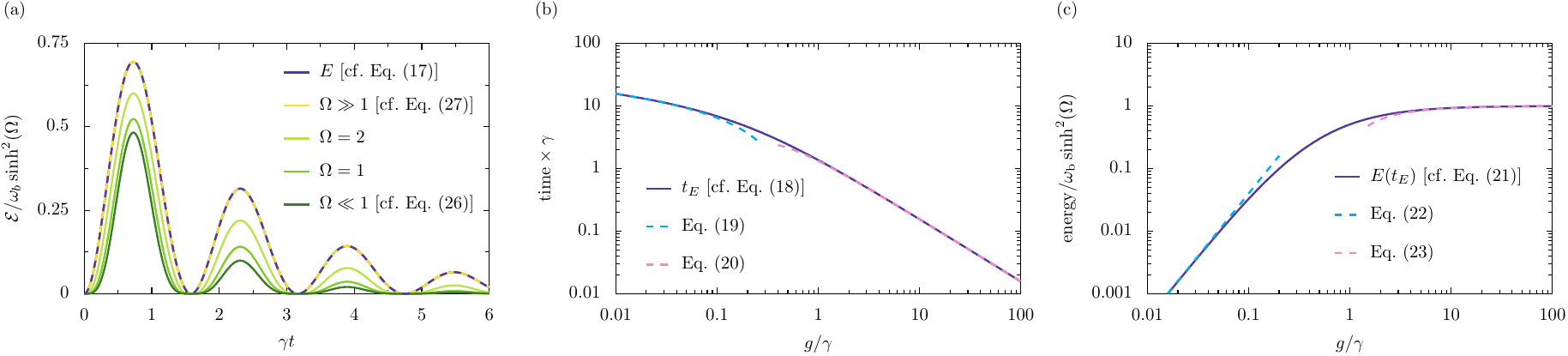}
 \caption{ \textbf{Energetics of the quantum battery.} Panel (a): the ergotropy $\mathcal{E}$ of the quantum battery in units of $\omega_b \sinh^2(\Omega)$, where $\omega_b$ is the energy level spacing and $\Omega$ is the strength of the driving pulse, as a function of time (in units of the inverse decay rate $1/\gamma$) [cf. the definitions of Eq.~\eqref{eq:fgddsfdfsfgfgd}--Eq.~\eqref{eq:svgverfeee}, along with the results of Eq.~\eqref{eq:sadscdasrfewr} and Eq.~\eqref{eq:fdvsdfvtr}]). We consider the case when the coupling strength $g = 2 \gamma$, and increasing large values of $\Omega$ are marked with increasingly light (dark green to yellow) lines. Purple line: the stored energy $E$ [cf. Eq.~\eqref{eq:sdfsfdsfdsf}]. Yellow line: the strong driving approximation of Eq.~\eqref{eq:sdfvdfvhghhh}. Dark green line: the weak driving approximation of Eq.~\eqref{eq:sadsasdadfssdfsdcdasrfewr}. Panel (b): a log-log plot of the time $t_E$ taken (in units of $1/\gamma$) to reach the maximum in stored energy $E$, as a function of the coupling-to-dissipation ratio $g/\gamma$ [cf. Eq.~\eqref{eq:sdfsdfs}]. Dashed cyan line: the weak coupling approximation of Eq.~\eqref{eq:sddsfsdffgsdf}. Dashed pink line: the strong coupling approximation of Eq.~\eqref{eq:vscesec}. Panel (c): a log-log plot of the maximum stored energy $E(t_E)$, in  units of $\omega_b \sinh^2(\Omega)$, as a function of $g/\gamma$ [cf. Eq.~\eqref{eq:sdfsfdfghdsasdafhsfdsf}]. Dashed cyan line: the weak coupling approximation of Eq.~\eqref{eq:adsdascadw}. Dashed pink line: the strong coupling approximation of Eq.~\eqref{eq:sfdcf}.}
 \label{figure1}
\end{figure*}

\noindent \textbf{Results}\\
The equation of motion for the second moments of interest, for all times $t > 0$ after the action of the driving pulse appearing in Eq.~\eqref{eq:sfdcvfdddn}, is given by the Schrödinger-like equation
\begin{equation}
\label{eq:sdfsfd}
\mathrm{i} \partial_t \psi = \mathcal{H} \psi,
\end{equation}
where the mean operators are contained within $\psi$, and where the four-dimensional dynamical matrix $\mathcal{H}$ are together defined by 
\begin{equation}
\label{eq:sdfssdsdfsdffd}
\psi = \begin{pmatrix}
\langle a^\dagger a \rangle \\
\langle b^\dagger b \rangle \\
\langle a^\dagger b \rangle \\
\langle b^\dagger a \rangle \\
\end{pmatrix},
\quad\quad
\setstackgap{L}{1.1\baselineskip}
\fixTABwidth{T}
\mathcal{H} = \parenMatrixstack{
    -\mathrm{i} \gamma & 0 & g & -g  \\
    0 & 0 & -g & g  \\
    g & -g & -\mathrm{i} \frac{\gamma}{2} & 0 \\
    -g & g & 0 & -\mathrm{i} \frac{\gamma}{2}
}.
\end{equation}
The formal solution of Eq.~\eqref{eq:sdfsfd} leads to the following expressions for the average populations $\langle a^\dagger a \rangle$ and $\langle b^\dagger b \rangle$ of the battery charger and battery holder respectively,
\begin{align}
\label{eq:sdsdfsdffsf}
\langle a^\dagger a \rangle &= C \left[ \tfrac{1}{2} \left( \tfrac{g}{G} \right)^2 + \tfrac{8g^2-\gamma^2}{16 G^2} \cos \left( 2 G t \right) - \tfrac{\gamma}{4 G} \sin \left( 2 G t \right) \right] \mathrm{e}^{-\frac{\gamma t}{2}}, \nonumber \\
\langle b^\dagger b \rangle &= C \left( \tfrac{g}{G} \right)^2 \sin^2 \left( G t \right) \mathrm{e}^{-\frac{\gamma t}{2}}, 
\end{align}
which gives rise to the renormalized coupling strength $G$, which is defined as
\begin{equation}
\label{eq:sdfsdfsd}
G = \sqrt{ g^2 - \left( \tfrac{\gamma}{4} \right)^2 },
\quad\quad\quad
\Gamma = \sqrt{  \left( \tfrac{\gamma}{4}  \right)^2 - g^2 },
\end{equation}
and where we also introduced the related renormalized decay rate $\Gamma$ for use later on. The solutions of Eq.~\eqref{eq:sdsdfsdffsf} are given in terms of an unknown constant prefactor $C$, which can be determined by the boundary conditions of the problem as defined by the parametric drive. Exactly at $t= 0$ the quadratic pulse of Eq.~\eqref{eq:sfdcvfdddn} acts, which couples a wider variety of moments such that Eq.~\eqref{eq:sdfsfd} becomes ten-dimensional (as shown in the Supplemental Material~\cite{SupplementalMaterial}). Integrating this generalized equation of motion, using the integrating factor technique for example, reveals that the $C$ prefactor is simply
\begin{equation}
\label{eq:asdasd}
C = \sinh^2 \left( \Omega \right).
\end{equation}
The consequence of the quadratic drive of Eq.~\eqref{eq:sfdcvfdddn} has been to introduce a degree of quantum squeezing~\cite{Walls1983, Loudon1987} into the quantum battery system. This has manifested in the hyperbolic enhancement factor $C$ of Eq.~\eqref{eq:asdasd}, where the dimensionless drive strength $\Omega$ resembles a kind of squeezing parameter (as is discussed in the Supplemental Material~\cite{SupplementalMaterial}). Importantly, such a striking hyperbolic enhancement as $C$ is not seen in linearly driven models due to the absence of squeezing in that more standard case~\cite{Ukhtary2024}.

Most notably, the mean populations of Eq.~\eqref{eq:sdsdfsdffsf} also exhibit an exceptional point when $g = g_{\mathrm{EP}}$, where we introduce the key value [cf. Eq.~\eqref{eq:sdfsdfsd}]
\begin{equation}
\label{eq:sdfsdfgdfgfdfsd}
g_{\mathrm{EP}} = \tfrac{\gamma}{4},
\end{equation}
which marks where two of the eigenvalues and eigenvectors associated with Eq.~\eqref{eq:sdfssdsdfsdffd} coalesce, in a type of spectral singularity~\cite{Berry2003, Miri2019, Downing2021, Fox2023}. The exceptional point of Eq.~\eqref{eq:sdfsdfgdfgfdfsd} demarcates the borderland between oscillatory population cycles and non-oscillatory population dynamics in the quantum battery. This drastic change in the dynamics is noticeable from the stored energy $E$, which follows from Eq.~\eqref{eq:fgdgfgd} and Eq.~\eqref{eq:sdsdfsdffsf} as 
\begin{equation}
\label{eq:sdfsfdsfdsf}
E = \begin{cases}
  \omega_b \sinh^2 \left( \Omega \right) \left( \tfrac{g}{\Gamma} \right)^2 \sinh^2 \left( \Gamma t \right) \mathrm{e}^{-\frac{\gamma t}{2}},  & g < g_{\mathrm{EP}}, \\
    \omega_b \sinh^2 \left( \Omega \right) \left( \frac{\gamma t}{4} \right)^2 \mathrm{e}^{-\frac{\gamma t}{2}},  & g = g_{\mathrm{EP}}, \\
 \omega_b \sinh^2 \left( \Omega \right) \left( \tfrac{g}{G} \right)^2 \sin^2 \left( G t \right) \mathrm{e}^{-\frac{\gamma t}{2}}, & g > g_{\mathrm{EP}}.
\end{cases}
\end{equation}
In particular, there is an unusual quadratic dependence exactly at the exceptional point, inbetween the more standard hyperbolic and trigonometric dynamics. There is always an exponential decay of the stored energy $E$ with the time constant $2/\gamma$ due to the lossy battery charger, as depicted in Fig.~\ref{figure0}. Most importantly, Eq.~\eqref{eq:sdfsfdsfdsf} demonstrates a hyperbolic enhancement in energy thanks to the $C$ factor of Eq.~\eqref{eq:asdasd}, making the pulsed quadratic battery an interesting proposition for a quantum storage device exploiting squeezing (as is further discussed in the Supplemental Material~\cite{SupplementalMaterial}). We plot the dynamical energy $E$ of Eq.~\eqref{eq:sdfsfdsfdsf}, in units of $\omega_b \sinh^2(\Omega)$ and for the example case with the coupling $g = 2 \gamma$, as the purple line in Fig.~\ref{figure1}~(a). This graph displays some hallmarks of the proposed quantum battery model in the desirable $g > g_{\mathrm{EP}}$ regime, including exponentially damped oscillations and exact energetic zeroes at certain times when the condition $t = n \pi/G$ (where $n$ is an integer) is fulfilled.

Clearly from the purple line (marking the stored energy $E$) in Fig.~\ref{figure1}~(a), the temporal position of the first energetic peak -- which is the global maximum in the energy $E$ -- is of great importance. Indeed, the knowledge of this energetic maximum allows one to find the optimal charging time $t_E$, where $E( t_E ) = \mathrm{max}_t \{ E (t) \}$. The turning points of Eq.~\eqref{eq:sdfsfdsfdsf} suggest that the optimal charging times $t_E$ are given by
\begin{equation}
\label{eq:sdfsdfs}
t_E = \begin{cases}
  \frac{\arctanh \left( \frac{4 \Gamma}{\gamma} \right)}{\Gamma},  & g < g_{\mathrm{EP}}, \\
  \frac{1}{g_{\mathrm{EP}}},  & g = g_{\mathrm{EP}}, \\
 \frac{\arctan \left( \frac{4 G}{\gamma} \right)}{G}, & g > g_{\mathrm{EP}},
\end{cases}
\end{equation}
which is plotted as the purple line in Fig.~\ref{figure1}~(b), as a function of the crucial coupling-to-dissipation ratio $g/\gamma$. The asymptotics of Eq.~\eqref{eq:sdfsdfs} at small and large ratios of $g/\gamma$ makes the scaling of the optimal time $t_E$ even more explicit as
\begin{align}
\label{eq:sddsfsdffgsdf}
\lim_{g \ll g_{\mathrm{EP}}} t_E &= \frac{4}{\gamma} \ln \left( \tfrac{\gamma}{2 g} \right), \\
\lim_{g \gg g_{\mathrm{EP}}} t_E &= \frac{\pi}{2g} - \frac{\gamma}{4g^2}, \label{eq:vscesec}
\end{align}
which are marked with the dashed cyan and dashed pink lines respectively in Fig.~\ref{figure1}~(b). Intuitively, with weaker couplings $g$ the optimal charging time $t_E$ increases -- and in fact it eventually becomes logarithmically divergent [cf. Eq.~\eqref{eq:sddsfsdffgsdf}]. Conversely, stronger couplings $g$ lead to quicker optimal times $t_E$, which rapidly approaches the inverse-$g$ dissipationless result of $t_E = \pi/2g$ [which itself comes from the first maximum of the sinusoidal function, cf. Eq.~\eqref{eq:vscesec}].

The energetic maximums $E ( t_E )$ themselves are determinable directly from Eq.~\eqref{eq:sdfsfdsfdsf} with Eq.~\eqref{eq:sdfsdfs}, leading to the compact results
\begin{equation}
\label{eq:sdfsfdfghdsasdafhsfdsf}
E \left( t_E \right) = \begin{cases}
  \omega_b \sinh^2 \left( \Omega \right) \mathrm{e}^{-\frac{ \gamma }{2\Gamma} \arccoth \left( \frac{\gamma}{4\Gamma} \right)},  & g < g_{\mathrm{EP}}, \\
    \omega_b \sinh^2 \left( \Omega \right) \mathrm{e}^{-2},  & g = g_{\mathrm{EP}}, \\
 \omega_b \sinh^2 \left( \Omega \right)  \mathrm{e}^{-\frac{ \gamma }{2G} \arccot \left( \frac{\gamma}{4G} \right)}, & g > g_{\mathrm{EP}},
\end{cases}
\end{equation}
where the case exactly at the exceptional point ($g = g_{\mathrm{EP}}$) sees the dimensionless number $1/\mathrm{e}^{2} \simeq 0.135$ arise. This optimal energy $E ( t_E )$ is plotted as the the purple line in Fig.~\ref{figure1}~(c) as a function of the reduced coupling strength $g/\gamma$. The limiting behaviour of this optimal energy follows from Eq.~\eqref{eq:sdfsfdfghdsasdafhsfdsf} as
\begin{align}
\label{eq:adsdascadw}
\lim_{g \ll g_{\mathrm{EP}}} E \left( t_E \right) &= \omega_b \sinh^2 \left( \Omega \right) \left(  \tfrac{2 g}{\gamma} \right)^2, \\
\lim_{g \gg g_{\mathrm{EP}}} E \left( t_E \right) &= \omega_b \sinh^2 \left( \Omega \right) \left( 1 - \tfrac{\pi \gamma}{4 g} \right), \label{eq:sfdcf}
\end{align}
which captures the quadratic in $g$ decrease in optimal energy at very small couplings where $g \ll \gamma$ [cf. Eq.~\eqref{eq:adsdascadw}], and defines the energetic upper bound of $\omega_b \sinh^2 ( \Omega )$, which is approached inverse-linearly in $g$ at large couplings where $g \gg \gamma$ [cf. Eq.~\eqref{eq:sfdcf}]. These asymptotics are represented with the dashed cyan and dashed pink lines respectively in Fig.~\ref{figure1}~(c), completing the basic energetic analysis of the proposed pulsed quantum battery.

In the discharging phase, when the battery charger is disconnected from the battery holder, the maximal amount of work which can be extracted from the quantum battery (via unitary operations) is governed by the ergotopy $\mathcal{E}$~\cite{Allahverdyan2004, Alicki2013}. For the case of the pulsed quantum battery, this important measure follows directly from the definitions of Eq.~\eqref{eq:fgddsfdfsfgfgd}--Eq.~\eqref{eq:svgverfeee}. Pleasingly, the results of Eq.~\eqref{eq:sdsdfsdffsf} and Eq.~\eqref{eq:asdasd}, as well as auxiliary results presented in the Supplemental Material~\cite{SupplementalMaterial}, allow for an explicit expression for the influential quantity $\mathcal{D}$ to be determined [cf. Eq.~\eqref{eq:svgverfeee}], via 
\begin{align}
\label{eq:sadscdasrfewr}
\mathcal{D} &= 1 + 4 \sinh^2 \left( \Omega \right) \mathcal{P} \left( 1 - \mathcal{P} \right), \\
\mathcal{P} & = \left( \tfrac{g}{G} \right)^2 \sin^2 \left( G t \right) \mathrm{e}^{-\frac{\gamma t}{2}}. \label{eq:fdvsdfvtr}
\end{align}
These brief formulae allow for the dynamical ergotropy $\mathcal{E}$ to calculated analytically, the results of which are shown in Fig.~\ref{figure1}~(a), where increasing large values of the drive $\Omega$ are marked with increasingly light (from dark green to light green to yellow) lines. The functional behaviour of the ergotopy $\mathcal{E}$ is highly reminiscent of the stored energy $E$ (purple line in the panel), essentially being damped sinusoidal-like oscillations. In the two limiting cases of very weak and very strong drives, the ergotropy $\mathcal{E}$ is captured by the exact asymptotics
\begin{align}
\label{eq:sadsasdadfssdfsdcdasrfewr}
\lim_{\Omega \ll 1} \mathcal{E}  &= \omega_b \sinh^2 \left( \Omega \right) \mathcal{P}^2, \\
\lim_{\Omega \gg 1} \mathcal{E}  &=  \omega_b \sinh^2 \left( \Omega \right) \mathcal{P},  \label{eq:sdfvdfvhghhh}
\end{align}
where the dynamic function $\mathcal{P}$ was just introduced in Eq.~\eqref{eq:fdvsdfvtr}. When $\Omega \ll 1$, there is a significant difference between the stored energy $E$ (purple line) and the ergotropy $\mathcal{E}$ (dark green line) as shown in Fig.~\ref{figure1}~(a), signifying the inefficiency of this weak driving regime [cf. Eq.~\eqref{eq:sadsasdadfssdfsdcdasrfewr}]. However, when $\Omega \gg 1$ the ergotropy $\mathcal{E}$ (yellow line) swiftly approaches the stored energy $E$ (purple line) exactly [cf. Eq.~\eqref{eq:sdfsfdsfdsf} and Eq.~\eqref{eq:sdfvdfvhghhh}]. Hence the pulsed quadratic battery exhibits a remarkable strong driving regime with two highly desirable measures of performance: (i) it is essentially perfectly efficient since $\mathcal{E} \simeq E$, and (ii) it is able to store an abundance of energy thanks to the hyperbolic enhancement carried by the all-important $C$ factor [cf. Eq.~\eqref{eq:asdasd}].
\\


\noindent \textbf{Discussion}\\
In conclusion, we have suggested a quantum continuous variable model of a quantum battery, which features a hyperbolic enhancement of its stored energy (due to its quadratic pulsed drive inducing quantum squeezing). This remarkable feature acts to combat the unavoidable energy loss suffered by the battery charger. Importantly, the ergotropy is essentially equal to the stored energy for strong enough driving pulses, revealing the efficient nature of the proposed quantum battery in the discharging phase. Our analytic theory also allows for compact expressions for the optimal charging times and optimal energetics of this high-performing quantum battery. The underlying idea of exploiting squeezing in order to boost the achievements of the quantum battery may also have some significance for wider efforts surrounding quantum technological advancements in the generation, transfer and storage of energy~\cite{Deutsch2020, Auffeves2022}. 
\\


\noindent \textbf{Acknowledgments}\\
\textit{Funding}: CAD is supported by the Royal Society via a University Research Fellowship (URF\slash R1\slash 201158) and by Royal Society Enhanced Research Expenses which support MSU. CAD also gratefully acknowledges an Exeter-FAPESP SPRINT grant with the Universidade Federal de São Carlos (São Paulo, Brazil). \textit{Discussions}: CAD thanks R.~Bachelard, A. Cidrim, A. C. Santos and C. J. Villas-Boas for fruitful discussions and for their hospitality during his visits to UFSCar.
\\



\end{document}